\documentstyle[aps,prd]{revtex}

\begin{document}

\title{Expectation values of four-quark operators in pions
}
\author{E.G. Drukarev, M.G. Ryskin, V.A.~Sadovnikova,\\
Petersburg Nuclear Physics Institute\\
Gatchina, 188300 St. Petersburg, Russia\\
 and \\
Amand~Faessler\\
Institute of Theoretical Physics, University of Tuebingen,
Morgenstelle 14, D-72076, Tuebingen, Germany }
\date{today}
\maketitle

\begin{abstract}
The values of four-quark operators averaged over pions are expressed
through those averaged over vacuum. The specific values are obtained in
the framework of the factorization assumption. For the condensates of
the light quarks of the same flavour $\bar q\Gamma q\bar q\Gamma q$ the
scalar condensate is shown to be an order of magnitude larger than the
other ones. The condensates containing the strange quarks $\bar q q\bar
s s$ appear to be only about twice smaller than those of the light
quarks.  The degeneracy of the ground state in the Nambu--Jona--Lasinio
model is shown explicitly.
\end{abstract}

\section{Introduction}

The matrix elements of four-quark operators in hadronic states provide
information on the structure of the quark-antiquark sea in the hadrons
describing the correlations of $\bar q q$ pairs. Under certain reasonable
assumptions the expectation value of the scalar quark operator $\bar q q$
was found in \cite{1}--\cite{3} to be equal (up to normalization
factors) to the total number of quarks and antiquarks $n_h$ in the
hadron. The known expectation values of $\bar q q$ in nucleons and pions
\cite{4} correspond to $n_N\approx8$ and $n_\pi\approx12$, exceeding
the number of the valence quarks. This signals  the large numbers of
the sea quarks even at low energies.

Another application is the QCD sum rules in nuclear matter. In-medium
values of the four-quark condensates is an important ingredient of the
approach \cite{5,6}. In the gas approximation it is expressed through
the expectation value in the nucleons.

As it stands now, the only calculation of the four-quark condensate in
hadrons is that carried out by Celenza et~al. \cite{7}. In \cite{7} the
condensate in the nucleons was obtained in the framework of the Nambu and
Jona--Lasinio model. Some data on its value in nucleons were obtained by
Johnson and Kisslinger \cite{8} by considering the QCD sum rules for
nucleons and isobars.

Here we present calculation of the expectation values of the four-quark
operators in pions. Besides providing information on the structure of
pions, the analysis is the intermediate step in the calculation of the
expectation values in nucleons in the framework of  models where the
quark-antiquark sea is contained in the pion cloud. Some of the models
analyzed in \cite{9} as well as perturbative chiral quark model
\cite{10},  provide the examples.

The general form of the four-quark expectation value can be presented
as
\begin{equation}
Q_{12}\ =\ \langle\pi|\bar q T_1q\cdot\bar q T_2q|\pi\rangle
\end{equation}
with the subscripts of the matrices $T_{1,2}$ including spin, colour
and flavour. By using the reduction formula obtained by Lehmann,
Symanzik and Zimmerman (LSZ) \cite{11,12} we present the matrix
elements of Eq.(1) through the expectation values in vacuum
\begin{equation}
Q_{12}\ =\ \frac1{f^2_\pi}\sum_{i,j}\langle0|\bar q T_iq\cdot\bar q
T_jq|0\rangle
\end{equation}
with $f_\pi\approx93$~MeV being the pion decay constant. Further
calculations are carried out in the framework of a factorization hypothesis
for the vacuum expectation values. This hypothesis was first formulated
by Shifman et~al. \cite{13} and was advocated recently in \cite{14}. In
this approximation the expectation values of the light quark operators
are expressed through the well known value \cite{15} of the condensate
\begin{equation}
\langle0|\bar q q|0\rangle\ =\ -\ \frac{m^2_\pi f^2_\pi}{m_u+m_d}
\end{equation}
with $m_{\pi,u,d}$ standing for the masses of the pion and of the light
quarks. For example, if $"q"$ is $"u"$ or $"d"$ quark
\begin{equation}
\langle\pi|(\bar q q)^2|\pi\rangle\ \approx\ \frac{-2(\langle0|\bar q q|0
\rangle)^2}{f^2_\pi}
\end{equation}
for any state of the pion isotope triplet.

The negative value of the condensate corresponds to a domination of
"disconnected terms" when one of two $\bar q q$ pairs comes from the vacuum.
The $\bar q q$ pairs in the pion are found to be correlated strongly, with
the probability of the coexistence of two pairs being much smaller than in
the independent pairs picture.

The four-quarks interactions are presented in Nambu and Jona--Lasinio
model \cite{16}, emerging in its simplest SU(2) version.
The instanton induced t'Hooft interaction \cite{17} is often included
in the applications nowadays-see e.g.\cite{18}.  Being written in terms
of the current quarks, these interactions $V_{1,2}$ correspond to the
picture "before" the symmetry was spontaneously broken. Thus the result
of averaging over the pion states
$\langle\pi|V_1|\pi\rangle=\langle\pi|V_2|\pi\rangle=0$ is a reasonable
one. This means that introducing the interactions $V_{1,2}$ does not
change the energy of the ground state. Hence, the latter remains
degenerate until the spontaneous symmetry breaking occurs.

\section{General equations}

We define the expectation value of any operator $A$ in a hadron $h$ as
\begin{equation}
\langle h|\widehat A|h\rangle\ =\ \left\langle h|\int
d^3x\left[\widehat A(x)-\langle0 |\widehat A(x)|0\rangle\right]
|h\right\rangle
\end{equation}
with the
vector of state $\langle h|$ normalized as $\langle h(k)\mid
h(p)\rangle=2p_0\delta({\bf k}-{\bf p})$. Of course, the expectation
value $\langle0|\widehat A(x)|0\rangle$ does not depend on $x$.
Thus the matrix element $\langle h|\widehat A|h\rangle$ is
defined as the excess of the density of the operator $A$ over the
vacuum value, integrated over the volume of the hadron.

Due to the partial conservation of axial current (PCAC) the pion state
vector can be expressed through that of the vacuum (see, e.g.,
\cite{12})
\begin{equation}
|\pi^\alpha(x)\rangle\ =\ \frac1{\sqrt2\,f_\pi m^2_\pi}\
\partial_\mu J^\alpha_{\mu5}(x)|0\rangle
\end{equation}
with $\alpha$ being the isospin index while $J^\alpha_{\mu s}$ is the
axial current of the light quarks with the corresponding quantum
numbers:
\begin{eqnarray}
&& J^-_{\mu5}(x)\ =\sum_c\bar d^c(x)\gamma_\mu\gamma_5u^c(x); \quad
J^+_{\mu5}(x)\ =\sum_c\bar u^c(x)\gamma_\mu\gamma_5d^c(x)\ ; \nonumber\\
&& J^0_{\mu5}(x)\ =\ \sum_c\frac{\bar u^c(x)\gamma_\mu\gamma_5u^c(x)
-\bar d^c(x)\gamma_\mu\gamma_5d^c(x)}{\sqrt2}
\end{eqnarray}
with $"c"$ being the colour index.

By applying Eq.(5) for both $\langle\pi|$ and $|\pi\rangle$ states in
the matrix element $\langle\pi|A|\pi\rangle$ of any operator $A$ one
finds the reduction LSZ formula \cite{11,12}
\begin{equation}
\langle\pi^\alpha|\widehat A|\pi^\alpha\rangle\ =\
\frac1{f^2_\pi}\ \langle0|B^\alpha|0\rangle
\end{equation}
with
\begin{equation}
B^\alpha=\ \frac{1}{2 V}\int d^3xdy_0dz_0\delta(x_0-y_0)\delta(z_0-x_0)
\left[\bar Q^\alpha_5(z_0), [Q^\alpha_5(y_0), A(x)]\right].
\end{equation}
Here $V$ is the normalization volume, the commutator $[X, Y]=XY-YX$,
while
\begin{equation}
Q^\alpha_5(y_0)\ =\ \int d^3yJ^\alpha_{05}(y)
\end{equation}
is the axial charge, corresponding to the current $J^\alpha_{\mu5}$.

Note that application of Eq.(9) to the quark scalar operator $A=\bar u
u+\bar d d$ yields \cite{4}
\begin{equation}
\langle\pi|\bar u u+\bar d d|\pi\rangle\ =\ \frac{2m^2_\pi}{m_u+m_d}
\end{equation}
with the proper behavior in the chiral limit. It was shown in
\cite{19} that the dimensionless value, identified with the total
number of quarks and antiquarks
\begin{equation}
n_\pi\ =\ \frac{\langle\pi|\bar u u+\bar d d|\pi\rangle}{2m_\pi}
\end{equation}
can be obtained by using the pion vector of state $|\tilde\pi\rangle$
with normalization $\langle\tilde\pi(k)|
\tilde\pi(p)\rangle=\delta({\bf k}-{\bf p})$.

\section{The four-quark operators}

The general form of the operator, containing the four quark fields $q$
of a specific flavour is $A=q^a_\alpha\bar q\,^b_\beta
q^c_\gamma\bar q\,^d_\delta$ with $\alpha,\beta,\gamma,\delta$ and
$a,b,c,d$ being Lorentz and colour indices. In the applications the
product of the fields is time-ordered. Thus, considering the product of
the field operators at the same space-time point we treat it as the
limit of a time-ordered product.

Each of the products of the two-quark fields can be presented as
\begin{equation}
q^a_\alpha\bar q^b_\beta\ =\ -\frac1{12}\sum_X\bar q\Gamma^Xq
\Gamma^X_{\alpha\beta}\delta_{ab}-\frac1{64}\sum_{X,\rho}\bar q\Gamma^X
\lambda^\rho q\Gamma^X_{\alpha\beta}\lambda^\rho_{ab}
\end{equation}
with $\lambda^\rho$ $(\rho=1\ldots8)$ standing for the standard
Gell-Mann matrices normalized by the relation $\rm
Sp\,\lambda^\rho\lambda^\tau=2\delta^{\rho\tau}$. The 16 basic Dirac
$4\times4$ matrices are
\begin{equation}
\Gamma^1=I\ , \quad \Gamma^2_\mu=\gamma_\mu\ , \quad
\Gamma^3=\gamma_5\ , \Gamma^4_\mu=\gamma_\mu\gamma_5\ , \quad
\Gamma^5_k=\frac12(\gamma_\mu\gamma_\nu-\gamma_\nu\gamma_\mu)
\end{equation}
with $\mu=1\ldots4$, $k=1\ldots6.$

Thus it is sufficient to study the condensates of the form
\begin{equation}
S_{XY}\ =\ \langle h|\sum_{a,b}\bar q\,^a\Gamma^Xq^a\cdot\bar q\,^b \Gamma^Y
q^b|h\rangle
\end{equation}
and
\begin{equation}
R_{XY}\ =\ \langle h|\sum_{a,a',b,b',\rho}\bar q\,^a\Gamma^X
\lambda^\rho_{aa'}q^{a'}\cdot\bar q\,^b\Gamma^Y\lambda^\rho_{bb'} q^{b'}|h
\rangle
\end{equation}
with the matrices $\Gamma(\lambda)$ acting on Lorentz (colour) indices.

For the expectation values of the operators, which are antisymmetric in
the colour variables
\begin{equation}
 U_{XY}\ =\ \langle h|\sum_{a,a',b,b'}\bar
q\,^a\Gamma^Xq^{a'}\cdot\bar q\,^b\Gamma^Yq^{b'} |h\rangle\
(\delta_{aa'}\cdot\delta_{bb'}-\delta_{ab'}\cdot\delta_{ba'})
\end{equation}
we find by using the properties of Gell-Mann matrices
\begin{equation}
U_{XY}\ =\ \frac23\ S_{XY}-\frac12\ R_{XY}
\end{equation}
with $S_{XY}$ and $R_{XY}$ being defined by Eqs. (15), (16).

The condensates determined by Eqs. (15), (16) contain the quarks
of the same flavour. For the case of different flavours of quarks:
$u,d(u,d,s)$ one can present similar covariant expressions replacing
$\Gamma^{X,Y}$ by $T^{X,Y}_i=\Gamma^{X,Y}\cdot t^i$ with $t^i$ acting
on the isospin indices.

\section{Expectation values of the light quark operators}

Here we calculate the expectation values of the 4-quark operators which
include $"u"$ and $"d"$ quarks only. The commutators which enter Eq.(8)
can be calculated by using the relation \cite{16}
\begin{equation}
\left\{q^a_\alpha(x), \bar q\ ^b_\beta(y)\right\}\ =\
(\gamma_0)_{\alpha\beta}\delta_{ab}\delta({\bf x}-{\bf y})
\end{equation}
(with $\{X, Y\}=XY+YX)$ at $x_0=y_0$, while the anticommutators between
all the other quark operators turn to zero. Evaluation of the
right-hand side (rhs) of Eq.(9) thus leads to the form
\begin{equation}
\langle0|B^\alpha|0\rangle\ =\ \frac1V\ \langle0|\int d^3xF^\alpha(x)|0
\rangle
\end{equation}
with $F^\alpha(x)$ being the product of four quark operators. Since the
vacuum is uniform, the density $F^\alpha$ does not depend on $x$.
Hence, we can put
\begin{equation}
B^\alpha\ =\ F^\alpha(0)\ .
\end{equation}

 To illustrate how it works we present the calculation of the operator
$B$, corresponding to the expectation value $\langle\pi^-|\bar u\Gamma
u\cdot\bar u\Gamma u|\pi^-\rangle$ with $\Gamma=\Gamma^X=\Gamma^Y$. The
internal commutator in the rhs of Eq.(9) is
\begin{equation}
\kappa\ =\ \int d^3y\sum_a\left[\bar d\,^a(y)\gamma_0\gamma_5u^a(y)\ ,
\ \bar u(x)\Gamma u(x)\bar u(x)\Gamma u(x)\right].
\end{equation}
Presenting explicitly
$\bar d\,^a\gamma_0\gamma_5u^a=\sum_{\gamma,\alpha}\bar d\,^a_\gamma(
\gamma_0\gamma_5)_{\gamma_\alpha}u^a_\alpha$ and $\bar u\Gamma
u=\sum_{\beta,\delta,b}\bar u\,^b_\beta\Gamma_{\beta\delta}u^b_\delta$ we
employ the fact that the anticommutator of the fields $u^a_\alpha$ and
$\bar u\,^b_\beta$ is given by Eq.(19) while all the others ones are
zero. Using also the relation $\{\gamma_0, \gamma_5\}=0$ we obtain
\begin{equation}
\kappa\ =\ -\left\{\bar d(x)\gamma_5\Gamma u(x)\ ,\ \bar u(x)\Gamma
u(x)\right\}.
\end{equation}
After similar calculation of the commutator
$[\bar Q_5^-,\kappa]$ we find
\begin{equation}
B^-\left((\bar u\Gamma
u)^2\right)\ =\ -\bar u\Gamma u\left(\bar u\Gamma u +\bar d\gamma_5
\Gamma\gamma_5d\right)-\bar d\gamma_5\Gamma u\bar u\Gamma\gamma_5d\ .
\end{equation}
Here we introduced notation $B^\alpha(A)$ for the operator $B$ related
to operator $A$ by Eq.(9).

One can obtain in the same way
\begin{equation}
B^+\left((\bar u\Gamma u)^2\right)\ =
B^-\left((\bar u\Gamma u)^2\right)\
\end{equation}
and
\begin{equation}
 B^0\left((\bar u\Gamma u)^2\right)\ =\ -\bar u\Gamma u\left(\bar
u\Gamma u +\bar u\gamma_5\Gamma\gamma_5u\right)-\frac12\bar
u\{\gamma_5, \Gamma\}u \cdot \bar u\{\gamma_5 ,\Gamma\}u\ .
\end{equation}

For the mixed operator $A=\bar u\Gamma u\bar d\Gamma d$ we find
\begin{equation}
B^{\pm}\left(\bar u\Gamma u\bar d\Gamma d\right)\ =\ -
 \bar u\Gamma u\bar d\Gamma d-\frac12\left(\bar u\Gamma u\bar
u\gamma_5\Gamma\gamma_5 u +\bar u\gamma_5\Gamma d\bar d\gamma_5\Gamma
u+(u\leftrightarrow d)\right),
\end{equation}
while
\begin{equation}
 B^0\left(\bar u\Gamma u\bar d\Gamma d\right)\ =\ -\bar u\Gamma u\bar
d\Gamma d -\frac12\left(\bar u\Gamma u\bar
d\gamma_5\Gamma\gamma_5d+\bar d\Gamma d \bar
u\gamma_5\Gamma\gamma_5u-\bar u\{\gamma_5,\Gamma\}u\,\bar d\{\gamma_5 ,
\Gamma\} d\right).
\end{equation}
Of course, the isotope invariance provides the relations $B^+\left((\bar d
d)^2\right)=B^-\left((\bar u u)^2\right)$, etc.

For the important special case $\Gamma=I$ Eqs. (24)--(28) take the form
\begin{eqnarray}
&& B^{\pm}\left((\bar u u)^2\right)\ =\ -(\bar u u)^2-\bar u u\bar d d
-\bar d\gamma_5u\bar u\gamma_5d \\
&& B^0\left((\bar u u)^2\right)\ =\ -2\left((\bar u u)^2
+(\bar u\gamma_5u)^2\right)
\end{eqnarray}
and
\begin{equation}
 B^{\pm}(\bar u u\bar d d)\ =\ -\frac12\left((\bar u u+\bar d
d)^2+2\bar u\gamma_5d\bar d \gamma_5u\right) ,
 \end{equation}
while
\begin{equation}
B^0(\bar u u\bar d d)\ =\ -2\left(\bar u u\bar d d-\bar u\gamma_5u\bar
d\gamma_5d\right).
\end{equation}

Note that Eqs. (24)--(32) are obtained for the operators $S_{XX}$
defined by Eq.(15), i.e. for the colourless diquarks. The operators
$B(R_{XX})$ with $R_{XX}$ defined by Eq.(16) are expressed by
the same Eqs. (25)--(28) with the matrices $\Gamma$ being replaced by
 $\widetilde\Gamma\,^\rho=\Gamma\lambda^\rho$ with further
summation over $\rho$.

By using Eqs. (9) and (18) one can obtain the operator $B(A)$ for the
four-quark operator of the general form
\begin{equation}
A\ =\ \sum_\rho\bar\Psi\widetilde\Gamma_i\,^\rho P_1\Psi\cdot\bar\Psi
\widetilde\Gamma\,^\rho_j P_2\Psi
\end{equation}
with $\Psi=\left({u \atop d}\right)$ being the quark spinor. The
operators $P_{1,2}$ are the projection operators in the isospin space.
Each of them is equal either to $P_+$ or to $P_-$ with
$P_{\pm}=(1\pm\tau_3)/2$. We defined
\begin{equation}
\widetilde\Gamma\,^\rho_k\ =\ \Gamma_k\lambda^\rho
\end{equation}
for any $4\times4$ matrix $\Gamma_k$ acting on Lorentz indices.

The matrix $\lambda^\rho\ (\rho=0\ldots8)$ is either the unit matrix
$(\rho=0)$  or one of the Gell-Mann colour matrices $(\rho=1,\ldots8)$.
For the operator $A$ defined by Eq.(33) we obtain
\begin{eqnarray}
&& B^\alpha=\
-\frac12\bar\Psi\left(\gamma_5\Gamma^\rho_j\tau_{\bar\alpha}P_2
+\Gamma_j^\rho\gamma_5P_2\tau_{\bar\alpha}\right)\Psi \nonumber\\
\times &&\bar\Psi\left(\gamma_5\Gamma_i^\rho\tau_\alpha
P_1+\Gamma^\rho_i\gamma_5P_1\tau_\alpha\right)\Psi-\frac12\,\bar\Psi
\Gamma_j^\rho P_2\Psi \nonumber\\
\times &&\left(\bar\Psi_1\Gamma^\rho_i\tau_{\bar\alpha}\tau_\alpha
P_1\Psi+\bar\Psi\Gamma^\rho_iP_1\tau_\alpha\tau_{\bar\alpha}\Psi
\right)+(1\to2)
\end{eqnarray}
with $\alpha=+,-,0$ standing for the pion isospin indices. For the
specific cases Eq.(35) takes a much more simple form since some of the
terms vanish: $\tau_+P_+\Psi=\tau_-P_-\Psi=0;$
$P_-\tau_+\Psi=P_+\tau_-\Psi=0$.

\section{Four-quark operators containing heavier quarks}

Here we calculate the expectation values of the operators containing
heavier quarks. Calculations, similar to those described in the
previous Section, provide
\begin{equation}
B^{\pm}\left(\bar u\Gamma^Xu\bar\psi_i\Gamma^Y\psi_i\right)\ =\
-\frac12\left(\bar u\Gamma^Xu+\bar d\gamma_5\Gamma^X\gamma_5d\right)
\bar\psi_i\Gamma^Y\psi_i\ ,
\end{equation}
while
\begin{equation}
B^0\left(\bar q\Gamma^Xq\bar\psi_i\Gamma^Y\psi_i\right)\ =\
-\frac12\left(\bar q\Gamma^Xq+\bar q\gamma_5\Gamma^X\gamma_5q\right)
\bar\psi_i\Gamma^Y\psi_i\ .
\end{equation}
For the operator $A$ consisting of the heavy quarks only, e.g.
$A=\bar\psi_i\Gamma^X\psi_i\bar\psi_j\Gamma^Y\psi_j$, we find
immediately
\begin{equation}
B^\alpha(A)\ =\ 0\ .
\end{equation}

\section{Factorization approximation}

 Now we calculate the rhs of Eqs. (24)--(37) in the factorization
approximation \cite{13}. This means that the sum over the intermediate
states is assumed to be dominated by the vacuum state, i.e.
\begin{eqnarray}
&& \langle0|\bar q\,^i_{\alpha,a}q^j_{\beta,b}\bar q\,^k_{\gamma,c}
q^\ell_{\delta,d}|0\rangle\ =\ \frac1{12^2}\bigg(\delta_{ij}
\delta_{\alpha\beta}\delta_{ab}\delta_{k\ell}\delta_{\gamma\delta}
\delta_{cd}\ - \nonumber\\
- && \delta_{i\ell}\delta_{\alpha\delta}\delta_{ad}\delta_{jk}
\delta_{\beta\gamma}\delta_{bc}\bigg)\ \langle0|\bar q\,^iq^i|0\rangle\
\langle0|\bar q\,^jq^j|0\rangle
\end{eqnarray}
with the upper indices standing for the flavour. For the quarks of the
same flavour this provides for any $4\times4$ matrices
$\Gamma_r,\Gamma_s$
\begin{equation}
 \langle0|\bar q\widetilde\Gamma^\rho_rq\bar
q\widetilde\Gamma^\rho_sq|0\rangle\ =\ \frac1{12^2}\left(\mbox{Sp
}\widetilde\Gamma^\rho_r\mbox{ Sp } \widetilde\Gamma^\rho_s -\mbox{Sp
}\widetilde\Gamma^\rho_r\widetilde\Gamma\,^\rho_s\right)
(\langle0|\bar q q|0\rangle)^2
\end{equation}
with $\widetilde\Gamma^\rho_k$ defined by Eq.(34).
For $\rho=0$, when $\lambda^\rho$ is the unit matrix and
$\widetilde\Gamma^0_k=\Gamma_k$, Eq.(40) takes the form \begin{equation}
\langle0|\bar q\Gamma_rq\bar q\Gamma_sq|0\rangle\ =\
\frac1{16}(\mbox{Sp }\Gamma_r\mbox{ Sp }\Gamma_s-\frac13\mbox{Sp }\Gamma_r
\Gamma_s)\left\langle0|\bar q q|0\rangle\right)^2\ ,
\end{equation}
while
\begin{equation}
 \langle0|\sum_{\rho=1}\bar q\widetilde\Gamma\,^\rho_rq\bar
q\widetilde\Gamma\,^\rho_sq|0 \rangle\ =\ -\frac19\mbox{ Sp
}\Gamma_r\Gamma_s\left(\langle0|\bar q q|0\rangle\right)^2\ .
\end{equation}

For the quarks with different flavours $i\neq j$ we find
\begin{equation}
\langle0|\bar q_i\Gamma_rq_i\bar q_j\Gamma_sq_j|0\rangle\ =\ \frac1{16}
\mbox{ Sp }\Gamma_r\mbox{ Sp } \Gamma_s\langle0|\bar q_iq_i|0\rangle
\langle0|\bar q_jq_j|0\rangle
\end{equation}
and
\begin{equation}
\langle0|\bar q_i\Gamma_rq_j\bar q_j\Gamma_sq_i|0\rangle\ =\
-\frac1{3\cdot16}\mbox{ Sp }\Gamma_r\Gamma_s\langle0|\bar q_iq_i
|0\rangle\,\langle0|\bar q_jq_j|0\rangle\ ,
\end{equation}
while
\begin{equation}
 \langle0|\sum_{\rho=1}\bar q_i\widetilde\Gamma^\rho_rq_j\bar
q_j\widetilde\Gamma\,^\rho_s q_i|0\rangle\ =\ -\frac19\mbox{ Sp
}\Gamma_r\Gamma_s\langle0|\bar q_iq_i |0\rangle\,\langle0|\bar
q_jq_j|0\rangle\ ,
\end{equation}
which is true for $i=j$ as well, and
\begin{equation}
\langle0|\sum_{\rho=1}\bar q_i\widetilde\Gamma^\rho_rq_i
\bar q_j\widetilde\Gamma^\rho_s
q_j|0\rangle\ =\ 0\ .
\end{equation}

Assuming also  isospin invariance, i.e. $\langle0|\bar u u|0\rangle =
\langle0|\bar d d|0\rangle=\langle0|\bar q q|0\rangle$ we find
\begin{eqnarray}
 &&\langle0|B^+\left((\bar u\Gamma u)^2\right)|0\rangle =\
-\frac18\left((\mbox{ Sp }\Gamma )^2-\frac13\mbox{ Sp }\Gamma^2\right)
\left(\langle0|\bar q q|0\rangle \right)^2\\
 && \langle0|B^-\left((\bar u\Gamma u)^2\right)|0\rangle =\
 \langle 0 |B^+\left((\bar u\Gamma u)^2\right)|0\rangle
\end{eqnarray}
while
\begin{equation}
 \langle0|B^0\left((\bar u\Gamma u)^2\right)|0\rangle =\
-\frac18\left((\mbox{ Sp }\Gamma)^2 -\frac13\mbox{ Sp
}\Gamma^2-\frac13\mbox{ Sp }\Gamma\gamma_5\Gamma \gamma_5+(\mbox{Sp
}\gamma_5\Gamma)^2\right) .
\left(\langle0|\bar q q|0\rangle \right)^2\\
\end{equation}
For the mixed operator $A=\bar u\Gamma u\bar d\Gamma d$ we obtain
\begin{equation}
\langle0|B^{\pm}|0\rangle =\ -\ \frac18\left((\mbox{Sp
}\Gamma)^2-\frac13\mbox{Sp }\Gamma \gamma_5\Gamma\gamma_5\right)
\left(\langle0|\bar q q|0\rangle \right)^2\\
\end{equation}
and
\begin{equation}
 \langle0|B^0|0\rangle =\ -\ \frac18\left((\mbox{Sp
}\Gamma)^2-(\mbox{Sp }\gamma_5 \Gamma)^2\right)\ .
\left(\langle0|\bar q q|0\rangle \right)^2\\
\end{equation}

We present also the result for the operators containing the coloured
diquarks. For the operator $\sum_\rho(\bar u\Gamma\lambda^\rho u)^2$ they
are
\begin{eqnarray}
&& \langle0|B^+|0\rangle =\ \frac29\mbox{ Sp }\Gamma^2(\langle0|\bar q
q|0\rangle)^2\\
&& \langle0|B^-|0\rangle =\ \frac29\mbox{ Sp
}\Gamma^2 (\langle0|\bar q q|0\rangle)^2\\
&& \langle0|B^0|0\rangle =\ \frac29\left(\mbox{ Sp }\Gamma^2+\mbox{ Sp
}\Gamma \gamma_5\Gamma\gamma_5\right)(\langle0|\bar q q|0\rangle)^2\ ,
\end{eqnarray}
while for $A=\sum_\rho\bar u\Gamma\lambda^\rho u\bar
d\Gamma\lambda^\rho d$
\begin{equation}
\langle0|B^{\pm}|0\rangle =\ \frac29\mbox{ Sp
}\Gamma\gamma_5\Gamma\gamma_5\ (\langle0|\bar q q|0\rangle)^2\ ; \quad
\langle0|B^0|0\rangle =\ 0\ .
\end{equation}
Equations (47)--(55) are true for any $4\times4$ matrix $\Gamma$. If
$\Gamma=\Gamma^X$ with $\Gamma^X$, being one of the basic matrices
defined by Eq.(14), the unit matrix $\Gamma=\Gamma^1$ is the only one
for which $\mbox{ Sp }\Gamma\neq0$. Thus the scalar condensate appears
to be numerically larger than the other condensates of the form
$A=(\bar q\Gamma^Xq)^2$. The values are
\begin{eqnarray}
\langle0|B^{\pm}\left((\bar q q)^2\right)|0\rangle =\ -\frac{22}{12}
(\langle0|\bar q q|0\rangle)^2 \\
\langle0|B^0\left((\bar q q)^2\right)|0\rangle =\
-\frac{20}{12}(\langle0|\bar q q|0\rangle)^2,
\end{eqnarray}
while for $A=\bar u u\bar d d$
\begin{eqnarray}
&& \langle0|B^{\pm}|0\rangle =\ -\frac{22}{12}\ (\langle0|\bar q
q|0\rangle)^2 \\
&& \langle0|B^0|0\rangle =\ -2\,(\langle0|\bar q q|0\rangle)^2\ .
\end{eqnarray}
Deviations of the coefficients in the rhs of Eqs (56)-(58) manifest
themselves in the units of the characteristic factor 1/12-see Eq
(41). Also, for the operator $A=\sum_\rho(\bar u\lambda^\rho u)^2$
\begin{equation}
\langle0|B^\pm|0\rangle=\frac89\ (\langle0|\bar q q|0\rangle)^2; \quad
\langle0|B^0|0\rangle=\frac{16}9\ (\langle0|\bar q q|0\rangle)^2\ ,
\end{equation}
while for $A=\sum_\rho\bar u\lambda^\rho u\bar d\lambda^\rho d$
\begin{equation}
\langle0|B^{\pm}|0\rangle =\ \frac89\ (\langle0|\bar q q|0\rangle)^2\
; \quad \langle0|B^0|0\rangle=0\ .
\end{equation}

For the expectation values of the operators containing heavier quarks
we obtain for the operator $A=\bar q\Gamma^Xq\bar\psi_i\Gamma^Y\psi_i$
\begin{equation}
\langle0|B|0\rangle=-\frac1{32}\left(\mbox{Sp }\Gamma^X\mbox{ Sp
}\Gamma^Y+ \mbox{ Sp }\Gamma^X\mbox{ Sp
}\gamma_5\Gamma^Y\gamma_5\right) \langle0|\bar q q|0\rangle\
\langle0|\bar\psi_i\psi_i|0\rangle\ ,
\end{equation}
while $\langle0|B|0\rangle=0$ for the operators containing the
coloured diquarks.

\section{Summary}

We solved the problem of expressing the expectation values of
four-quark operators in pions through those in vacuum. The further
specific calculations were carried out in the framework of the
factorization hypothesis.

For the scalar operators of the light quarks the expectation values
turn out to be negative. This can be understood in the following way.
Note that the four-quark condensates take the form of the superposition
of the products of the two-quark operators $\eta_{1,2}$, i.e.,
$A=\eta_1\eta_2$. For $\eta_1=\eta_2=\eta$ Eq.(5) can be presented as
\begin{equation}
\langle h|\eta^2|h\rangle \ =\ 2\langle0|\eta|0\rangle\, \langle
h|\eta|h\rangle +\langle h|\int d^3x(\eta(x)-\langle0|\eta|0\rangle
)^2|h\rangle\ ,
\end{equation}
if the factorization approximation is assumed.

The first term in the rhs of Eq.(63) describes the "disconnected
contribution" with one of $\bar q q$ pairs coming from vacuum. For the
expectation values of the operators $(\bar q q)^2$, i.e. $\eta=\bar q
q$, the first term in the rhs of Eq.(63) is exactly the rhs of Eq.(4).
Thus the expectation values of operators $(\bar q q)^2$ are dominated
by "disconnected terms". The same refers to the operator $A=\bar u
u\bar d d$. For the operators $\eta$ of the form $\bar q\Gamma^Xq$ with
$X\neq1$ (see Eq.(14)) the "disconnected terms " vanish since the
scalar condensate is the only one with a nonzero vacuum expectation
value.

For the scalar operators the second term of rhs of Eq.(63) describes
the correlation between $\bar q q$ pairs inside the pion. It corresponds
to deviations of the rhs of Eqs. (56)--(59) from that of Eq.(4). If these
pairs would have been independent, we would expect the estimation
\begin{equation}
 \langle0|B|0\rangle -(-2) \cdot \frac{\langle0|\bar q q|0\rangle
^2}{f^2_\pi} \approx \langle \pi |\bar q q|\pi\rangle \cdot
\frac{n^q_\pi}{V_\pi}
\end{equation}
to be true. Here $n^q_\pi$ being
the number of $\bar q q$ pairs of the certain flavour (see Eq.(12)),
while $V_\pi$ is the pion volume. The latter can be obtained since the
pion radius is known to be \cite{20}
\begin{equation}
r_\pi\ =\ \frac{\sqrt3}{2\pi f_\pi}\ .
\end{equation}
However, the value obtained from Eqs. (56)-(59) appears
to be much smaller than the rhs of Eq(64). This means that the pairs
of light quarks are correlated strongly and that the probability of
finding two pairs at the same point is much smaller than in the
approximation of independent pairs.

As one can see from the formulae of Sec.5, for the condensates of the
same flavour $\bar q\Gamma^Xq\bar q\Gamma^Xq$ those with $X=1$ are about 10
 times larger than the other values of $X$. In other words, the value
of the scalar condensate is about 10 times larger than the values of
the vector, pseudoscalar, pseudovector and tensor condensates. This is
because $\Gamma^1$ is the only matrix with Sp$\,\Gamma^X\neq0$. Since
for each of the matrices $\Gamma^X$ defined by Eq.(14) $|\rm
Sp\,(\Gamma^X)^2|=4$, the condensates with $X\neq1$ contain the
numerically small factor $\frac1{3\cdot4}=\frac1{12}$. For the
condensates with different flavours $\bar u\Gamma^Xu\bar d\Gamma^Xd$
the scalar condensate dominates for the charged pions. For the neutral
pions the pseudoscalar condensate is as large as the scalar condensate.

We obtain also the values for the condensates containing heavier
quarks. Assuming $\langle 0|\bar s s|0\rangle\approx \langle0|\bar q q|0
\rangle $ with $q$ standing for the $"u"$ or $"d"$ quark, we find
the expectation values $\langle \pi|\bar q q\bar s s|\pi\rangle $ to be only
about twice smaller than the value $\langle \pi|(\bar q q)^2|\pi\rangle$.
The expectation values, containing two pairs of heavier quarks ($s,c,$
etc.)  become zero.
 The expectation values of the four-quark operators in pions in the
terms describing the four-quark interactions in NJL model are zero.
This is the expected result corresponding to the degeneracy of the
ground state before the spontaneous symmetry breaking.

We thank M. Eides and T.~Gutsche for useful discussions and E.Gerjuoy
for reading the manuscript. Two of us (E. G. D.) and (V. A. S.) are
grateful for the hospitality of University of Tuebingen during
their visits. The work was supported by DFG grant
No.438~RUS~113/595/0-1 and by RFBR grant No.00-02-16853.

\end{document}